\documentclass{article}
\usepackage{graphicx}
\usepackage{amsmath}
\usepackage{amsfonts}
\usepackage{amssymb}

\begin{document}

\title{Noncommutative Cosmology}
\author{M. I. Beciu\\Dept. of Physics, Technical University\\B-d Lacul Tei 124, sect. 2 Bucharest -Romania \\e-mail: beciu@hidro.utcb.ro}
\maketitle
\begin{abstract}
We show that the large number coincidence can be interpreted as giving the
filling factor in a Landau problem. The analogy with the Landau problem leads
to noncommutativity between the gravitational and matter degrees of freedom.
We present a toy model that supports this view. We present, also, some of the
physical consequences this noncommutativity implies like a different insight
into the semiclassical approximation of quantum gravity and a different
tackling of the cosmological constant problem.
\end{abstract}

The large number coincidence [1] fascinated [2], [3] and still fascinates [4]
many people. The aim of focusing upon it was to find the reasons of the link
between micro and macro cosmos the coincidence seems to reveal.

Our aim here is to show that the large number coincidence hints to a
commutation relation. Unlike the noncommutativity in spacetime assumed in many
recent papers [5], the noncommutativity suggested by the large number
coincidence is in superspace, between the gravitational and matter degrees of freedom.

The large number coincidence, expressed in its most usual form%

\begin{equation}
\left(  \frac{\hbar^{2}H}{G\cdot c}\right)  ^{1/3}\sim m \label{this}%
\end{equation}

connects the Hubble parameter $H=a/a$ with the fundamental constants $h,c,G$
and a typical hadron mass $m$. For power law cosmologies, the scale factor
varies like $a(t)\propto t^{\alpha}$ and one can replace the Hubble constant
with the horizon distance $R_{H}$%

\begin{equation}
R_{H}=c\cdot a(t)\int_{0}^{t}\frac{dt^{\prime}}{d(t^{\prime})}=\frac
{ct}{1-\alpha}\quad;\quad H=\frac{\alpha}{t}\quad;\quad H=\frac{\alpha
\ c}{(1-\alpha)R_{H}}%
\end{equation}

Replacing also for mass in (1), the associated Compton wavelength $\lambda$,
we get%

\begin{equation}
\frac{\hbar^{2}\cdot\alpha}{(1-\alpha)R_{H}\cdot G}=\frac{h^{3}\cdot\alpha
}{\lambda\cdot c_{3}}%
\end{equation}

or, after a little rearrangement%

\begin{equation}
\left(  \frac{R_{H}}{\lambda}\right)  ^{3}=O(...)\frac{R_{H}\cdot\lambda}{2\pi
l_{p}^{2}}%
\end{equation}

where $l_{P}$ is the Planck length and $O(...)$ is a numerical factor,
something between $0.01$ and $1$, depending on the type of cosmology we
consider $(\alpha)$, the definition of the Planck length, the hadron mass we
choose. Using a reasoning dating back to Eddington, we can say the LHS of (4)
represents, roughly, the number of particles in a universe with horizon
$R_{H}$. If a single particle was present in the universe, its extension would
be of same order as the extension of that universe $\lambda/R_{H}=\delta
R_{H}/R_{H}\propto1$. When $N$ particles are present, the statistical argument
gives $\lambda/R_{H}=\delta R_{H}/R_{H}\propto N^{-1/2}$, wherefrom
$N\propto\left(  R_{H}/\lambda\right)  ^{2}$.

The RHS is more interesting; it resembles the degeneracy of an energy level in
the Landau problem. We remind that, in the Landau problem of a charged
particle placed a strong magnetic field and confined to a two dimensional
plane, say $X,Y$ the degeneracy of an energy level and the filling factor are, respectively%

\begin{equation}
D=\frac{L_{x}L_{y}}{2\pi l_{B}^{2}}\quad;\quad\nu=\frac{N}{D}%
\end{equation}

where $l_{B}=(hc/e\ B)^{1/2}$ is the magnetic length and $L_{x},L_{y}$ are the
sizes of the bidimensional sample. In the analogy above, the sizes are $R_{H}$
and $\lambda$, and the role of the magnetic length is played by the Planck
length. Note that, in this case, the analogs of the coordinates $X,Y$ must be
two degrees of freedom, one associated to a gravitational degree of freedom,
say the scale factor $a$ or a function of it, and the other one, to a matter
degree of freedom, say a scalar field. Relation (4) also points to a filling
factor for Universe $\nu\leq1$, i.e. the Universe accomodates (almost) the
maximum allowed number of particles.

It is a well known fact [6] that the dynamics of a charged particle in 2D
subject to a strong constant magnetic field $B$ is equivalent to the dynamics
of the same particle, with no magnetic field present but confined to a
noncommutative plane with the algebra of coordinates%

\begin{equation}
\left[  x,y\right]  =i\frac{\hbar c}{eB}%
\end{equation}

The equivalence holds true in the limit of very strong magnetic field when the
magnetic length $l_{B}$ (and here the Planck length) is much smaller than any
other length scale occurring in the problem and when the Hilbert space is
truncated to the lowest Landau level $(n=1)$; it is probably, also true for
higher Landau levels, when the Hilbert space is truncated to the first n
Landau levels and the RHS of relation (6) is multiplied by a factor $n$ [7].

When we draw the analogy between the degeneracy in the Landau problem and the
RHS of (4) we do not suppose the existence of a mysterious magnetic field;
rather, we think relation (4) hints to a nonvanishing commutator between two
degrees of freedom, yet to be specified, one associated to the gravitation,
the other, to matter.

We might expect that the degrees of freedom alluded above, the analogs of the
noncommuting coordinates, shall be the variables in a minisuperspace model;
described by Wheeler-DeWitt equation. Consequently, Wheeler-DeWitt equation
must bear some resemblances to Schrodinger equation in Landau problem. This
does not happen, at first sight. First, the very appearence of Schrodinger
equation in Landau problem depends on the chosen gauge (but the expression of
degenaracy is gauge invariant). Then, despite its name of Schrodinger equation
for quantum gravity, Wheeler-DeWitt equation is hyperbolic, the signature in
the general case is -+++++ unlike the stationary Schrodinger equation. That is
the main obstruction to any comparison.

Let us consider, however, the following simple minisuperspace model of a
spatially homogenous and isotropic universe with metric%

\begin{equation}
ds^{2}=\sigma^{2}\left(  dt^{2}-a^{2}d\Omega_{3}^{2}\right)
\end{equation}

where $d\Omega_{3}^{2}$ is the metric on a 3-sphere of unit radius. The only
gravitational degree of freedom, $a$, which for later convenience we denote by
$X_{2}\equiv a$, is made dimensionless by choosing $\sigma=(2/3\pi)^{1/2}%
l_{P}$ The matter degree of freedom is represented by a conformally invariant
scalar field $\phi$. The Wheeler-DeWitt equation is then [8]%

\begin{equation}
\frac{1}{2}\left[  -\frac{\partial^{2}}{\partial X_{1}^{2}}+X_{1}^{2}%
+\frac{\partial^{2}}{\partial X_{2}^{2}}-X_{2}^{2}\right]  \Psi\left(
X_{1},X_{2}\right)  =0
\end{equation}

where $X_{1}=\pi^{3/2}3^{1/2}\phi/l_{P}$. The Hamiltonian is of the form%

\begin{equation}
H=H_{1}-H_{2}\quad;\quad H_{i}=\frac{P_{i}^{2}+X_{i}^{2}}{2}\ ,\ i=1,2
\end{equation}

and it constituted object of special interest for 't Hooft in a recent series
of papers [9]. Eventually, we can transform the above Hamiltonian into%

\begin{equation}
H=yp_{x}-xp_{y}%
\end{equation}

using the transformations%

\[
P_{1}=\frac{1}{\sqrt{2}}\left(  p_{x}+y\right)  \quad;\quad P_{2}=\frac
{1}{\sqrt{2}}\left(  x+p_{y}\right)
\]%

\begin{equation}
X_{1}=\frac{1}{\sqrt{2}}\left(  x-p_{y}\right)  \quad;\quad X_{2}=\frac
{1}{\sqrt{2}}y-p_{x}%
\end{equation}

Neither the Hamiltonian (9) nor the Hamiltonian (10) is bounded from below. We
invoke now the procedure advocated by 't Hooft for Hamiltonians (9) or (10).
The lack of lower bound for the Hamiltonian (10) is cured changing to a
positive definite function $\rho^{2}$ that commute with (9)%

\begin{equation}
\left[  \rho^{2},H\right]  =0
\end{equation}

so that%

\begin{equation}
H_{1,2}=\frac{1}{4\rho^{2}}\left(  \rho^{2}\pm H\right)  ^{2}\quad;\quad
H=H_{1}-H_{2}%
\end{equation}

To get a lower bounded Hamiltonian, one imposes as a constraint, motivated by
information loss%

\begin{equation}
H_{2}\left|  \Psi\right\rangle =0
\end{equation}

whence%

\begin{equation}
H\rightarrow H_{1}\quad\rightarrow\rho^{2}\geq0
\end{equation}

A positive function that fulfils the above conditions is%

\begin{equation}
\rho^{2}=\frac{1}{2}\left(  x^{2}+y^{2}\right)
\end{equation}

A minimal requirement for 't Hooft prescription to work is that the
Hamiltonian (16) shall generate the same equations of motion as (10). It is
easy to check that this is possible iff the following bracket holds%

\begin{equation}
\left\{  x,y\right\}  =1
\end{equation}

To see more clearly the connection with the Landau problem we turn the bracket
into the commutator%

\begin{equation}
\lbrack x,y]=i
\end{equation}

The commutation relation can be implemented with the star Moyal product [10]
in momentum space and the Hamiltonian becomes%

\begin{equation}
\rho^{2}\ast\Psi=\frac{1}{2}\left(  x^{2}+y^{2}\right)  \ast\Psi=\frac{1}%
{2}\left(  \frac{P_{x}^{2}}{4}+\frac{P_{y}^{2}}{4}+x^{2}+y^{2}+p_{x}%
y-p_{y}x\right)  \Psi
\end{equation}

It represents exactly the Hamiltonian (in symmetric gauge) for a charged
particle of mass $\mu=2$ in constant magnetic field ($eB=4$). Using commutator
(18) and going through transformations (11) we get%

\begin{equation}
\left[  \phi,a\right]  =iC
\end{equation}

where, for this toy model, $C=2l_{p}/(3\pi^{3}a^{2})^{1/2}$. The relation
between 't Hooft procedure and Landau problem has also been proved by Banerjee
[11], in a slightly different manner and in a completely different context.

Can we go beyond this simple model equipped with a conformal scalar field? A
realistic superspace model involves, in general, interaction terms between $a$
and $\Phi$, and the two Hamiltonians do not decouple so nicely.

We can consider, however, the more realistic model of a homogenous isotropic
universe with a (nonconformal) scalar field $\phi$ with mass $\mu$. The
Wheeler-DeWitt equation is [12]%

\begin{equation}
\frac{1}{2}\left[  \frac{\partial^{2}}{\partial a^{2}}-a^{2}-\frac{1}{a^{2}%
}\frac{\partial^{2}}{\partial\Phi^{2}}+a^{4}m^{2}\Phi^{2}\right]  \Psi\left(
a,\phi\right)  =0
\end{equation}

where $\Phi=\sigma\phi$\ and $m=\sigma\mu$. As long as the mass of the field
is much smaller than the Planck mass $m\ll1$, the last term in (21) is
negligible small. Then, with the change of variables%

\begin{equation}
x=a\sinh\Phi\quad;\quad x=a\cosh\Phi
\end{equation}

equation (21) is brought to equation (8) to which the above analysis can be
applied. The commutation relation (20) will survive but with a different $C$.

Let us follow now the implications of a nonvanishing commutator between the
gravitational and matter degrees of freedom. An obvious consequence is the
fact that the problem of the singularity $a\rightarrow0$ is alleviated and
less worrisome. Due to the uncertainty relation%

\begin{equation}
\Delta a\Delta\phi\geq\frac{l_{p}}{\left(  3\pi^{3}\right)  ^{1/2}\left\langle
a\right\rangle }%
\end{equation}

the geometry becomes, as the singularity is approached, more and more fuzzy.

Another consequence concerns the so called semiclassical approximation in
quantum gravity. We recall that the semiclassical approximation consists in
treating classically the gravitational field while matter fields are treated
quantum mechanically. Relation (20) shows this procedure can be consistent;
one can \textit{not} quantize \textit{both} the gravitational and matter
fields at a time simply because they are not compatible observables.

The commutation relation sheds new light on the cosmological constant problem.
The cosmological constant in Einstein equations can be thought of as a purely
geometrical term $\Lambda$, proportional to the scalar curvature; let us call
it the geometrical cosmological constant $\Lambda_{g}$. A cosmological
constant can also occur due to matter from a stress tensor with the special
equation of state $p=-\rho$; let us call it the matter cosmological constant.
In terms of a scalar field $\phi$ the previous equation entails zero kinetic
energy and, since the field is constant, so is any arbitrary function of it,
in particular, the potential energy density $V(\phi)$. The cosmological term reads:%

\begin{equation}
\Lambda_{m}g_{\mu\ \nu}=-8\pi GTg_{\mu\ \nu}%
\end{equation}

where $T=V(\phi)/2$. The main point is that the two cosmological constants are
completely equivalent. There is no operational way to distinguish between
geometrical and matter cosmological constant at the classical level.%

\begin{equation}
\Lambda_{g}=\Lambda_{m}%
\end{equation}

On the other hand, when we compute the commutators by means of (20), we have%

\begin{equation}
\left[  \Lambda_{g}g_{\mu\ \nu},a\right]  =0\quad;\quad\left[  \Lambda
_{m}g_{\mu\ \nu},a\right]  =-8\pi GTg_{\mu\ \nu}\frac{dT}{d\phi}\left[
\phi,a\right]  \neq0
\end{equation}

Relations (25) and (26) are in clear contradiction. In physical terms, the
conflict is between the necessity of a strictly constant scalar field for the
cosmological constant and the forever fluctuating scalar field in (20). Put
differently, the tension is between the equivalence (25) and the broken
equivalence introduced by (20). A way out of these contradictions is to set to
zero in Einstein equations any term proportional to the metric. May be this
requirement does not solve completely the cosmological constant problem but it
forbids, for instance, the constant term occurring in a phase transition
governed by Higgs mechanism. The requirement above does not preclude the
existence of a scalar field with a small but nonvanishing kinetic energy; the
newly discovered accelerated expansion [13] could be driven by such a field
with negative pressure if $\left|  \ p\right|  <\rho$.

\bigskip

Both the empirical evidence, (the large number coincidence) and the model
above point to a nonnull commutator between matter and gravitational degrees
of freedom. At first sight the idea of a nonnull commutator between the matter
and gravitational degrees of freedom might seem crazy. We think it is crazy
enough to be true.

\bigskip

$\bigskip$

$\bigskip$

$\bigskip$

$\bigskip$

$\bigskip$

${\huge Appendix}$

We skipped in text some calculations. For completitude we give them here.

\textbf{1.} The Hamiltonian (16) with the bracket (17) engenders the same
equations of motion as the Hamiltonian (10).%

\[
H=yp_{x}-xp_{y}%
\]%

\[
a)\quad\frac{\partial H}{\partial p_{x}}=\overset{\cdot}{x}=y\quad;\quad
\frac{\partial H}{\partial p_{y}}=\overset{\cdot}{y}=x
\]%

\[
b)\quad\{x,y\}\quad;\quad\rho=\frac{1}{2}\left(  x^{2}+y^{2}\right)
\]%

\[
\overset{\cdot}{y}=\left\{  y,\rho\right\}  =\left\{  y,\frac{1}{2}\left(
x^{2}+y^{2}\right)  \right\}  =\left\{  y,x\right\}  x=-x
\]%

\[
\overset{\cdot}{x}=\left\{  x,\rho\right\}  =\left\{  x,\frac{1}{2}\left(
x^{2}+y^{2}\right)  \right\}  =\left\{  x,y\right\}  y=y
\]

\bigskip

\textbf{2.} A realization of commutation relation (18) is made by the star
$\left(  \ast\right)  $ product. The star $\left(  \ast\right)  $ product
leads to (19).

We define%

\[
\ast=\exp\left(  \frac{i}{2}\left(  \overleftarrow{\partial}_{x}%
\overrightarrow{\partial}_{y}-\overleftarrow{\partial}_{y}\overrightarrow
{\partial}_{x}\right)  \right)
\]

where $\overleftarrow{\partial}_{x}=\dfrac{\overleftarrow{\partial}}%
{\partial_{x}}$ acts at left etc.%

\[
x\ast y=x\left(  \exp\left(  \frac{i}{2}\left(  \overleftarrow{\partial}%
_{x}\overrightarrow{\partial}_{y}-\overleftarrow{\partial}_{y}\overrightarrow
{\partial}_{x}\right)  \right)  \right)  y=x\left(  1+\frac{i}{2}\left(
\overleftarrow{\partial}_{x}\overrightarrow{\partial}_{y}-\overleftarrow
{\partial}_{y}\overrightarrow{\partial}_{x}\right)  +...\right)  y=xy+\frac
{i}{2}%
\]%

\[
y\ast x=y\left(  1+\frac{i}{2}\left(  \overleftarrow{\partial}_{x}%
\overrightarrow{\partial}_{y}-\overleftarrow{\partial}_{y}\overrightarrow
{\partial}_{x}\right)  +...\right)  x=xy-\frac{i}{2}%
\]%

\[
\lbrack x,y]=x\ast y-y\ast x=i
\]%

\[
x^{2}\ast\Psi(x,y)=x^{2}\left(  1+\frac{i}{2}\left(  \overleftarrow{\partial
}_{x}\overrightarrow{\partial}_{y}-\overleftarrow{\partial}_{y}\overrightarrow
{\partial}_{x}\right)  +1+\frac{i^{2}}{8}\left(  \overleftarrow{\partial}%
_{x}\overrightarrow{\partial}_{y}-\overleftarrow{\partial}_{y}\overrightarrow
{\partial}_{x}\right)  ^{2}+...\right)  \Psi=
\]%

\[
=x^{2}\Psi+\frac{i}{2}2x\overrightarrow{\partial}_{y}\Psi+\frac{i^{2}}%
{8}x\overleftarrow{\partial}_{x}\overrightarrow{\partial}_{y}\overleftarrow
{\partial}_{x}\overrightarrow{\partial}_{y}\Psi=
\]%

\[
=x^{2}\Psi-xp_{y}+\frac{i^{2}}{4}\overrightarrow{\partial}_{y}\overrightarrow
{\partial}_{y}\Psi=x^{2}\Psi-xp_{y}+\frac{1}{4}p_{y}^{2}%
\]

where we took into account $i\partial_{y}=-p_{y}$.%

\[
\frac{1}{2}\left(  x^{2}+y^{2}\right)  \ast\Psi(x,y)=x^{2}+y^{2}-xp_{y}%
+yp_{x}+\frac{1}{4}p_{y}^{2}+\frac{1}{4}p_{x}^{2}\text{.}%
\]

\bigskip/

\textbf{3.} Equation (21) with the change of variables (22) is of the same
form as (8).%

\[
\left\{
\begin{array}
[c]{c}%
x=a\sinh\Phi\\
y=a\cosh\Phi
\end{array}
\right.
\]%

\[
\frac{\partial\Psi}{\partial a}=\frac{x}{a}\frac{\partial\Psi}{\partial
x}+\frac{y}{a}\frac{\partial\Psi}{\partial y}\quad;\quad\frac{\partial^{2}%
\Psi}{\partial a^{2}}=\frac{x}{a^{2}}\frac{\partial\Psi}{\partial x}+\frac
{y}{a^{2}}\frac{\partial\Psi}{\partial y}+\left(  \frac{x}{a}\right)
^{2}\frac{\partial^{2}\Psi}{\partial x^{2}}+\left(  \frac{y}{a}\right)
^{2}\frac{\partial^{2}\Psi}{y^{2}}%
\]%

\[
\frac{\partial\Psi}{\partial\Phi}=a\left(  \cosh\Phi\frac{\partial\Psi
}{\partial x}+\sinh\Phi\frac{\partial\Psi}{\partial y}\right)  \quad
;\quad\frac{\partial^{2}\Psi}{\partial\Phi^{2}}=x\frac{\partial\Psi}{\partial
x}+y\frac{\partial\Psi}{\partial y}+y^{2}\frac{\partial\Psi}{\partial x^{2}%
}+x^{2}\frac{\partial\Psi}{\partial y^{2}}%
\]%

\[
\frac{\partial^{2}\Psi}{\partial a^{2}}-\frac{1}{a^{2}}\frac{\partial^{2}\Psi
}{\partial\Phi^{2}}-a^{2}\Psi=\frac{\partial^{2}\Psi}{\partial y^{2}}%
-\frac{\partial^{2}\Psi}{\partial x^{2}}-y^{2}\Psi+x^{2}\Psi
\]
\end{document}